\documentclass[twoside,fleqn]{article}
\usepackage{espcrc2}
\usepackage{epsfig,psfrag,macros}
\usepackage{amssymb,amsmath}
\title{%
{
\vspace{-0.75cm}
\normalsize\hfill\parbox{21.0mm}{\raggedleft%
MS-TP-08-13
}}\\
Heavy quark masses from lattice QCD%
\thanks{Invited talk at the Workshop on \emph{${\rm e}^+{\rm e}^-$ Collisions 
from $\Phi$ to $\Psi$ (PHIPSI08)}, April 7--10, 2008, 
INFN Frascati, Italy.}
}

\author{%
Jochen Heitger\address{%
Westf\"alische Wilhelms-Universit\"at M\"unster, 
Institut f\"ur Theoretische Physik,\\
Wilhelm-Klemm-Strasse~9, D-48149 M\"unster, Germany}
}

\begin{document}

\begin{abstract}
I outline the basic strategies for the computation of charm and bottom
quark masses by means of lattice QCD, where particular emphasis is placed
on the non-perturbative renormalization of the effective theory for the
b-quark in heavy-light systems. 
A few selected results in the quenched approximation are reviewed, and
the current status of extending these calculations to QCD with dynamical 
quarks is summarized.
\end{abstract}

\maketitle
%
\section{Introduction}
\label{Sec_intro}
The fundamental parameters of QCD are the strong coupling constant and the
quark masses.
In particular, high-precision tests of the SM critically depend on the 
masses of the charm and bottom quarks, and the non-perturbative nature of 
lattice QCD, which allows to directly compute hadronic observables to match 
experimental inputs, is especially suited to determine them.

Their ab initio lattice calculation faces systematic errors, such as cutoff 
and finite-volume effects, too large dynamical light quark masses and the 
omission of some (or, in quenched QCD, even all) sea quark flavours. 
Among these, the problem of discretization errors proportional to powers of 
the bare quark mass is of utmost relevance in case of systems containing a 
heavy quark, when a Wilson-like fermion action is employed.
Requiring $a\mqh\ll 1$, it is then obvious that simulating a b-quark at its 
physical mass of $\approx 5\,\GeV$ in a space-time volume of
$L^4=(2\,\Fm)^4$ would demand $\gg 50$ lattice points per direction, 
an effort beyond the computing resources available today.

In this situation, an attractive possibility is to recourse to an 
\emph{effective theory} for the b-quark.
Extracting physical predictions from it, however, involves a matching to 
QCD, which becomes a severe source of uncertainty when performed 
perturbatively, for in the continuum limit results are power divergent.
Before coming to the effective theory framework to deal with heavy-light 
systems on the lattice (\Sect{Sec_npHQET}), let us consider the c-quark, 
where direct simulations are just doable.
\section{The charm quark's mass}
\label{Sec_Mc}
Physics in the region around the charm quark, which is by a factor of 
about 4 lighter than the b-quark, can be studied with conventional lattice
QCD methods, provided that --- assuming Wilson fermions here ---
the leading $\Or(a)$ cutoff effects have been eliminated non-perturbatively
and a range of lattice resolutions $a^{-1}\approx(2-4)\,\GeV$ is covered
in the simulations. 
This is nicely illustrated by the benchmark calculation~\cite{mcbar:RS02}
of the mass of the c-quark in the quenched approximation, which accounts
for all systematic errors except for the neglection of dynamical quark.
Following the strategy and techniques used to compute 
$\mS$~\cite{msbar:pap3}, and $\Or(a\mqc$) effects $\propto(\ba-\bp)$ being 
removed~\cite{impr:babp}, the renormalization group invariant (RGI) c-quark 
mass was calculated on 4 lattices with spacings $a=(0.1-0.05)\,\Fm$.
In addition to the definition of the (non-perturbatively) renormalized 
quark mass via the heavy-light PCAC relation, two further definitions were 
considered, differing by $\Or(a^2)$ discretization errors.
The agreement of all three definitions in the continuum limit found 
in~\cite{mcbar:RS02} thus provides a clear demonstration that on lattices of 
the chosen size (from $16^3\times 32$ to $32^3\times64$) one can control 
the discretization errors carefully by an extrapolation.

%
%
\begin{figure}[htb]
\centering
\psfrag{aaaaaaaa}[lc][l][1][0]{\begin{minipage}%
{\linewidth}\mbox{}\\[-7.0mm]\hspace*{-10mm}
\Huge$M_{\rm c}^{\rm (cs)}$
\end{minipage}}
\psfrag{b}[lc][l][1][0]{\begin{minipage}%
{\linewidth}\mbox{}\\[-3.0mm]\hspace*{-10mm}
\Huge$M_{\rm c}^{\rm (cc)}$
\end{minipage}}
\psfrag{c}[lc][l][1][0]{\begin{minipage}%
{\linewidth}\mbox{}\\[-5.0mm]\hspace*{-10mm}
\Huge$M_{\rm c}^{\rm (c)}$
\end{minipage}}
\psfrag{xlabel}[t][c][1][0]{\Huge$(a/r_0)^2$}
\psfrag{ylabel}[b][b][1][0]{\Huge$r_0M_{\rm c}$}
\epsfig{file=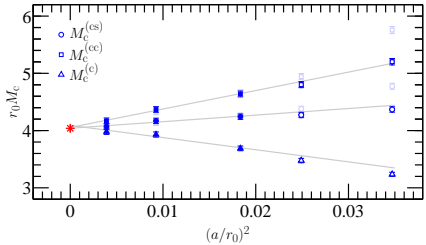,angle=-90,scale=0.225}
\vspace{-0.825cm}
\caption{\sl%
Continuum extrapolation of the RGI c-mass for $\nf=0$, 
covering $a=(0.1-0.03)\,\Fm$; see~\cite{fds:final} for details. 
The length scale $r_0=0.5\,\Fm$ is derived from the force between static 
quarks.
}\label{fig:Mc}
\vspace{-0.825cm}
\end{figure}
This is confirmed by including a further gauge field ensemble in a volume 
of $48^3\times 96$ with an even finer resolution of $a=0.03\,\Fm$, which was 
first generated in the context of \cite{thesis:andreasj} to investigate, 
within quenched QCD, the size of cutoff effects in the charm sector 
thoroughly also for other observables such as the $\Ds$-meson decay 
constant.
\Fig{fig:Mc} shows an update of the continuum extrapolation of the RGI
charm mass after a significant increase of the statistics compared 
to~\cite{mcbar:RS02,thesis:andreasj}, particularly for the smallest 
$a$~\cite{fds:final}.
The final result of this analysis, $\mcbMS(\mcb)=1268(24)\,\MeV$, 
is compatible with the earlier result of~\Ref{mcbar:RS02}.

However, from the figure it is also evident that a controlled assessment 
of the cutoff effects will not be possible for much heavier quarks.
Hence, for b-quarks other approaches must be devised, and one will be 
discussed in the next section.

Charm-quark mass computations in QCD with dynamical flavours are not yet 
in the stage such that a solid continuum limit extrapolation can be 
performed~\cite{mcharm:ukqcd,mcharm:etm_lat07}.
Since non-perturbative estimates for all the necessary improvement 
coefficients and renormalization factors in the two-flavour Wilson theory 
have become available now
\cite{impr:ca_nf2,impr:za_nf2,msbar:Nf2,impr:babp_nf2}, 
$\nf=2$ calculations of masses and matrix elements in the charm sector 
along the lines of the quenched studies are being started~\cite{mcharm:Nf2}.

For other recent determinations of $\mcbMS$ using QCD sum rules and 
current-current correlators in lattice and continuum QCD, 
see~\Refs{mcharm:karlsruhe,mcharm:hpqcd}.
\section{Non-perturbative HQET}
\label{Sec_npHQET}
Heavy Quark Effective Theory (HQET) at zero velocity on the 
lattice~\cite{stat:eichhill1} offers a reliable solution to the problem of 
dealing with the two disparate intrinsic scales encountered in heavy-light 
systems involving the b-quark, i.e., the lattice spacing $a$, which has to 
be much smaller than $1/m_{\rm b}$ to allow for a fine enough resolution of 
the states in question, and the linear extent $L$ of the lattice volume, 
which has to be large enough for finite-size effects to be under control.

Since the heavy quark mass ($\mb$) is much larger than the other scales 
such as its 3--momentum or $\lQCD\sim 500\,\MeV$, HQET relies upon a 
systematic expansion of the QCD action and correlation functions in inverse 
powers of the heavy quark mass around the static limit ($\mb\to\infty$).
The lattice HQET action, $S_{\rm HQET}$, at $\Or(1/\mb)$ reads:
\[
\hspace{-0.625cm}
a^4{\T \sum_x}\heavyb\left\{
D_{0}+\dmstat-\omkin\vecD^2-\omspin\vecsigma\vecB
\right\}\heavy\,,
\]
with $\heavy$ satisfying $P_+\heavy=\heavy$, $P_+={{1+\gamma_0}\over{2}}$,
and the parameters $\omega_{\rm kin}$ and $\omega_{\rm spin}$ being formally 
$\Or(1/\mb)$.
At leading order (static limit), where the heavy quark acts only as a 
static colour source and the light quarks are independent of the heavy 
quark's flavour and spin, the theory is expected to have $\sim 10\%$ 
precision, while this reduces to $\sim 1\%$ at $\Or(1/\mb)$ representing the 
interactions due to motion and spin of the heavy quark. 
As crucial advantage (e.g., over NRQCD), HQET treats the 
$1/\mb$--corrections to the static theory as space-time insertions in 
correlations functions, viz.
\bean
\hspace{-0.6775cm}
&&
\langle\op{}\rangle=\langle\op{}\rangle_\mrm{stat}+\\
\hspace{-0.6775cm}
&&
a^4\sum_x\{
\omkin\langle\op{}\Okin(x)\rangle_\mrm{stat}
+\omspin\langle\op{}\Ospin(x)\rangle_\mrm{stat}\}
\eean
for multi-local fields $O$, where $\langle\op{}\rangle_{\rm stat}$ denotes 
the expectation value in the static approximation and 
$\Okin$ and $\Ospin$ are given by $\heavyb\vecD^2\heavy$ and 
$\heavyb\vecsigma\vecB\heavy$.
In this way, HQET at a given order is (power-counting) renormalizable and
its continuum limit well defined, once the mass counterterm $\dmstat$ and
the coefficients $\omkin$ and $\omspin$ are fixed non-perturbatively by a 
matching to QCD.

Still, for lattice HQET and its numerical applications to lead to precise 
results with controlled systematic errors in practice, two shortcomings had 
to be left behind first.
\ben
\item
The exponential growth of the noise-to-signal ratio in static-light 
correlation functions, which can be overcome by a clever modification of the 
Eichten-Hill discretization of the static action~\cite{HQET:statprec}.
\item
As in HQET mixings among operators~of different dimensions occur, the
power-divergent additive mass renormalization $\dmstat\sim g_0^2/a$
already affects its leading order.
Unless HQET is renormalized non-perturbatively~\cite{Maiani:1992az}, this 
divergence --- and further ones $\sim g_0^2/a^{2}$ arising at 
$\Or(1/\mb)$ --- imply that the continuum limit does not exist owing to a 
remainder, which, at any finite perturbative 
order~\cite{mbstat:dm_MaSa,mbstat:dm_DirScor}, diverges as $a\to 0$.
A general solution to this theoretically serious problem was worked out and 
implemented for a determination of the b-quark's mass in the static and 
quenched approximations as a test case in~\Ref{HQET:pap1}.
It is based on a \emph{non-perturbative matching of HQET and QCD in 
finite volume}.
\een

In \Sect{Sec_npHQET_Mb} I review the HQET computation of the mass of the 
b-quark including the $\Or(1/\mb)$ terms for $\nf=0$~\cite{HQET:mb1m}, 
while \Sect{Sec_npHQET_Nf2} briefly summarizes the present status of the 
ongoing project to extend this to the more realistic case of two-flavour 
QCD~\cite{HQET:Nf2}.
A first application of the strategy of~\cite{HQET:pap1} to the computation 
of the B-meson decay constant can be found in~\cite{lat07:nicolas}.

A promising, not unrelated approach to lattice heavy-light systems derives 
from the step scaling method in the relativistic theory proposed 
in~\cite{mb:roma2}, where extrapolations in the heavy quark mass of step 
scaling functions (SSFs) as finite-size effects of proper observables, 
combined with SSFs calculated in HQET, turn into safer 
interpolations~\cite{mbfb:Nf0}.
\subsection{The b-quark mass in HQET at $\Or(\frac{1}{\mb})$}
\label{Sec_npHQET_Mb}
As pointed out in~\cite{reviews:NPRrainer_nara}, let us first note that 
in order not to spoil the asymptotic convergence of the series, the 
matching must be done non-perturbatively --- at least for the leading, 
static piece --- as soon as the $1/\mb$--corrections are included, since
as $\mb\to\infty$ the \emph{perturbative} truncation error from the 
matching coefficient of the static term becomes much larger than the power 
corrections $\sim\lQCD/\mb$ of the HQET expansion.

%
%
\begin{figure}[htb]
\centering
\epsfig{file=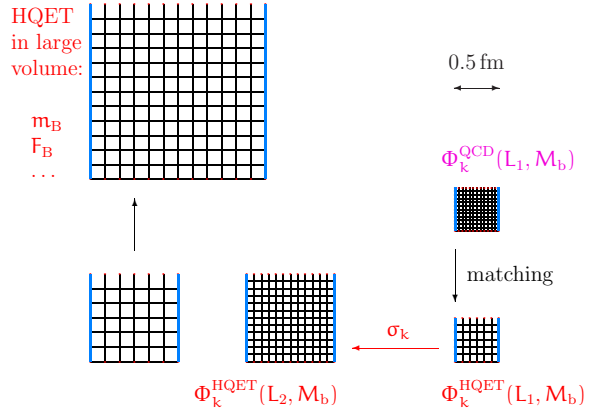,width=0.475\textwidth}
\vspace{-1.375cm}
\caption{\sl%
Idea of lattice HQET computations via a non-perturbative determination of 
HQET parameters from small-volume QCD simulations.
Arrows indicate steps to be repeated at smaller $a$ to reach a continuum 
limit.
(Drawing from~\cite{reviews:NPRrainer_nara}.)
}\label{fig:strat}
\vspace{-1.0cm}
\end{figure}
In the framework introduced in~\Ref{HQET:pap1}, matching and 
renormalization are performed simultaneously \emph{and} non-perturbatively.
The general strategy, illustrated in~\fig{fig:strat}, can be explained as
follows.
Starting from a finite volume with $L_1\approx 0.5\,\Fm$, one chooses 
lattice spacings $a$ sufficiently smaller than $1/\mb$ such that the 
b-quark propagates correctly up to controllable discretization errors of 
order $a^2$. 
The relation between the RGI and the bare mass in QCD being known, 
suitable finite-volume observables $\Phi_k(L_1,\Mb)$ can be calculated as a 
function of the RGI b-quark mass, $\Mb$, and extrapolated to the continuum 
limit. 
Next, the power-divergent subtractions are performed 
non-perturbatively by a set of matching conditions, in which the results
obtained for $\Phi_k$ are equated to their representation in HQET
(r.h.s.~of~\fig{fig:strat}).
At the same physical value of $L_1$ but for resolutions $L_1/a=\rmO(10)$, 
the previously computed heavy-quark mass dependence of $\Phi_k(L_1,\Mb)$ in 
finite-volume QCD may be exploited to determine the bare parameters of HQET 
for $a\approx(0.025-0.05)\,\Fm$.
In order to evolve the HQET observables to large volumes, where contact with 
experiments can be made, one also computes them at these lattice spacings in 
a larger volume, $L_2=2L_1$.
The resulting relation between $\Phi_k(L_1)$ and $\Phi_k(L_2)$ is encoded in
associated SSFs $\sigma_k$, as indicated in~\fig{fig:strat}.
Finally, the knowledge of $\Phi_k(L_2,\Mb)$ and employing resolutions 
$L_2/a=\rmO(10)$ fixes the bare parameters of the effective theory
for $a\approx(0.05-0.1)\,\Fm$ so that a connection to lattice spacings is 
established, where large-volume observables, such as the B-meson mass or 
decay constant, can be calculated (l.h.s.~of~\fig{fig:strat}). 
This sequence of steps yields an expression of $\mB$ (the physical input) as 
a function of $\Mb$ via the quark mass dependence of $\Phi_k(L_1,\Mb)$, 
which eventually is inverted to arrive at the desired value of the RGI 
b-mass within HQET.
The whole construction is such that the continuum limit can be taken for 
all pieces. 

More specifically, upon restricting to spin-averaged quantities to get rid 
of the contributions proportional to $\omspin$, the task is to fix 
$\dmstat$ and $\omkin$ non-perturbatively by performing a matching to QCD.
For sensible definitions of the required matching observables, 
$\Phi_1$ and $\Phi_2$, we work with the Schr\"odinger functional (SF), 
i.e.~QCD with Dirichlet boundary conditions in time and periodic ones in 
space (up to a phase $\theta$ for the fermions):
$\Phi_1^{\rm QCD}(L,\mh)$ exploits the sensitivity of SF correlators to 
$\theta$ and $\Phi_2^{\rm QCD}(L,\mh)\equiv L\Gamma_1(L,\mh)$, where 
$\Gamma_1$ is a finite-volume effective energy. 
When expanded in HQET\footnote{%
Here, $\dmstat=0$ in the action; its contribution is accounted for in the 
overall energy shift $\mhbare$ in HQET versus QCD.
}, 
$\Phi_1^{\rm HQET}(L)$ is given by $\omkin$ times a quantity defined in the 
effective theory (called $R_1^{\rm kin}(L,\theta,\theta')$), whereas
$\Phi_2^{\rm HQET}(L)$ is a function of $\omkin$ and 
$\mhbare=\dmstat+\mh$ involving two other HQET quantities, 
$\Gamma_1^{\rm stat}(L)$ and $\Gamma_1^{\rm kin}(L)$. 
According to the strategy sketched above, by equating 
$\Phi_k^{\rm QCD}(L_1,\mh)$ and $\Phi_k^{\rm HQET}(L_1)$ one can determine the 
bare parameters $\mhbare$ and $\omkin$ as functions of $\mh$ at the lattice 
spacings belonging to the volume $L_1^4$. 
To use the spin-averaged B-meson mass, $\mB^{\rm av}$, as phenomenological
input, the $\Phi_k$ are evolved to larger volumes through proper
SSFs~\cite{HQET:mb1m}, where the resulting $\Phi_k^{\rm HQET}(2L_1,\mh)$ still 
carry the dependence on $\mh$ inherited from the matching to QCD in $L_1^4$.
After 2 evolution steps (and taking continuum limits), linear extents of 
$\gtrsim 1.5\,\Fm$ are reached, and $\mhbare$ and $\omkin$, expressed in terms 
of SSFs, $\Phi_k^{\rm QCD}(L_1,\mh)$ as well as $R_1^{\rm kin}$, 
$\Gamma_1^{\rm stat}$ and $\Gamma_1^{\rm kin}$, are obtained 
--- again as functions of $\mh$.
Now, the b-quark mass is extracted by solving
\be
\mB^{\rm av}= 
E^{\rm stat}+\omkin(\mh)E^{\rm kin}+\mhbare(\mh)
\label{mastereq_E}
\ee
for $\mh$, with 
$E^{\rm stat}=\lim_{L\to\infty}\Gamma_1^{\rm stat}$ and
$E^{\rm kin}=
-\langle {\rm B}|a^3\sum_{\bf z}\Okin(0,\fat{z})|{\rm B}\rangle_{\rm stat}$.
All quantities entering \eq{mastereq_E} having a continuum limit either in 
QCD or HQET implies that all power divergences have been subtracted 
non-perturbatively.

%
%
\begin{figure}[htb]
\centering
\vspace{-0.75cm}
\epsfig{file=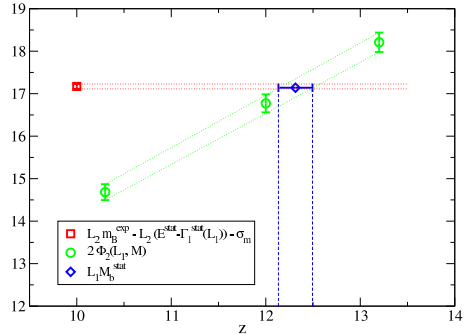,width=0.375\textwidth}
\vspace{-1.0cm}
\caption{\sl%
Graphical solution of \eq{mastereq_diff} in the quenched 
approximation~\cite{HQET:mb1m}.
The quantity used in the finite-volume matching step is 
$\Phi_2^{\rm QCD}(L_1,M)=L_1\Gamma_1(L_1,M)$, and $z\equiv L_1M$.
}\label{fig:Mb_stat}
\vspace{-0.75cm}
\end{figure}
In case of the leading-order, static approximation, where only $\mhbare$ 
needs to be determined, the small- and large-volume matching conditions
simplify to $\Gamma_1(L_1,\mh)=\Gamma_1^{\rm stat}(L_1)+\mhbare$
and $\mB^{\rm av}=E^{\rm stat}+\mhbare$, respectively.
To be able to replace $\mhbare$ in the latter by the former, we bridge the 
volume gap in two steps by inserting a SSF 
$\sigm(L_1)=2L_1[\Gamma_1^{\rm stat}(2L_1)-\Gamma_1^{\rm stat}(L_1)]$ and arrive
at the \emph{master equation}
\be
L_1\,[\,\mB^{\rm av}-(E^{\rm stat}-\Gamma_1^{\rm stat})\,]
-{\T \frac{\sigm(L_1)}{2}}=L_1\Gamma_1\,,
\label{mastereq_diff}
\ee
where $\Gamma_1=\Gamma_1(L_1,\mh)$ stems from $L_1^4$--QCD and any 
reference to bare parameters has finally disappeared.
Its graphical solution is reproduced in~\fig{fig:Mb_stat} and yields
$\Mb^{\rm stat}=6.806(79)\,\GeV$.

For the details on the (technically more involved) inclusion of the 
sub-leading $1/\mb$--effects, which exploits the freedom in choices for the 
angle $\theta$ as well as an alternative set of matching observables, 
I refer to~\Ref{HQET:mb1m}. 
Here, I just quote their final result $\mbbMS(\mbbar)=4.347(48)\,\GeV$
with the remark that, upon including the $1/\mb$--terms, differences among 
the static results w.r.t.~the matching condition chosen are gone, which 
signals practically negligible higher-order corrections.
\subsection{Status in two-flavour QCD}
\label{Sec_npHQET_Nf2}
The renormalization of HQET through the non-perturbative matching to 
$N_{\rm f}=2$ QCD in finite volume, to perform the power-divergent 
subtractions, is under way~\cite{HQET:Nf2}.
As an important prerequisite, the non-perturbative relation between the RGI 
and subtracted bare heavy quark mass calculated 
in~\cite{msbar:Nf2,impr:babp_nf2} enables to fix RGI heavy quark masses in 
the matching volume $L_1^4$:
\[
L_1M=
\zM(g_0)Z(g_0)\left(1+\bm(g_0)a\mqh\right)L_1\mqh\,.
\]
The extent $L_1$ is defined via a constant SF coupling, 
$\gbsq(L_1/2)=2.989$, and the PCAC masses of the dynamical light quarks are 
tuned to zero.

\begin{figure}[htb]
\centering
\vspace{-1.075cm}
\epsfig{file=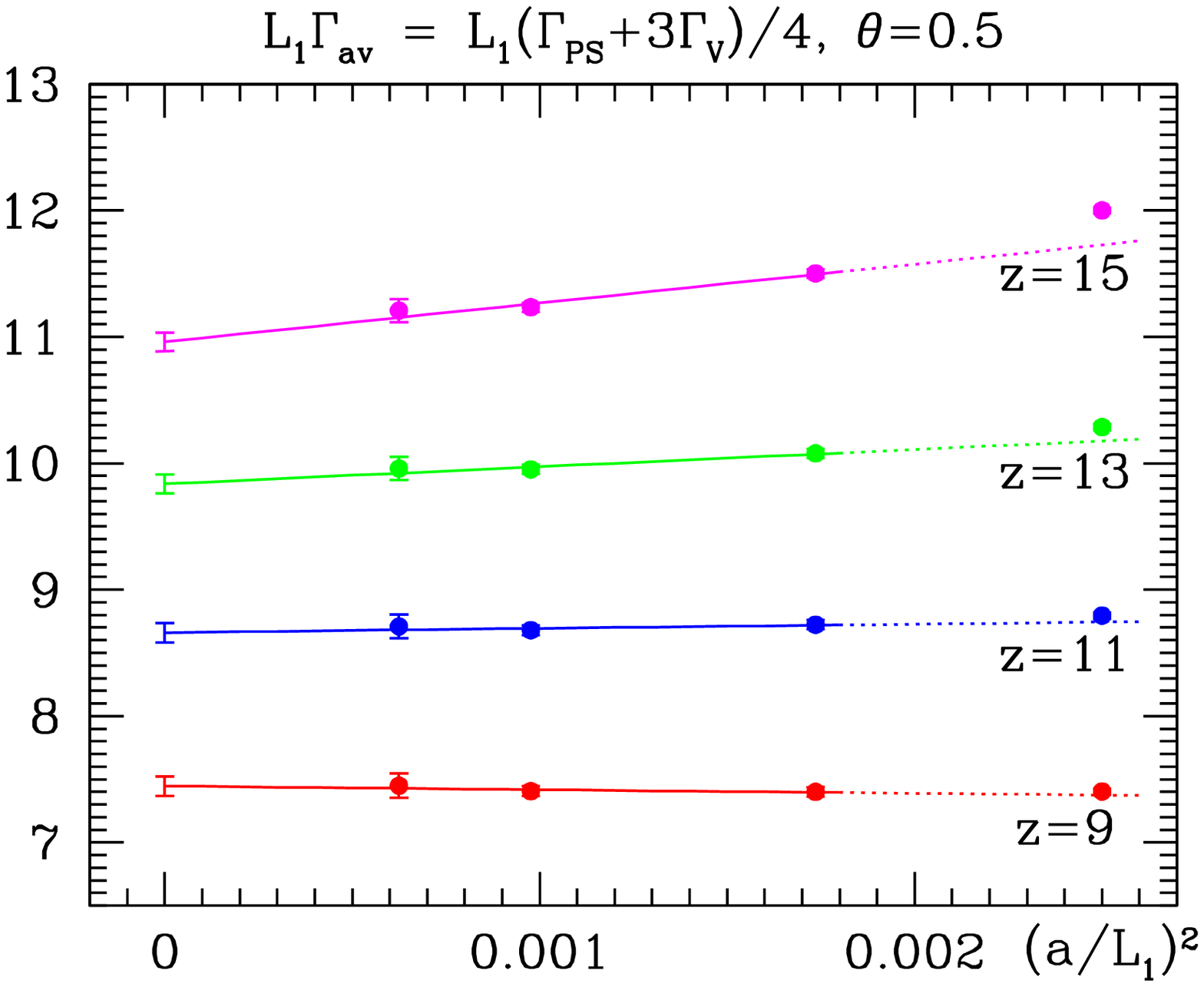,width=0.23075\textwidth}
\epsfig{file=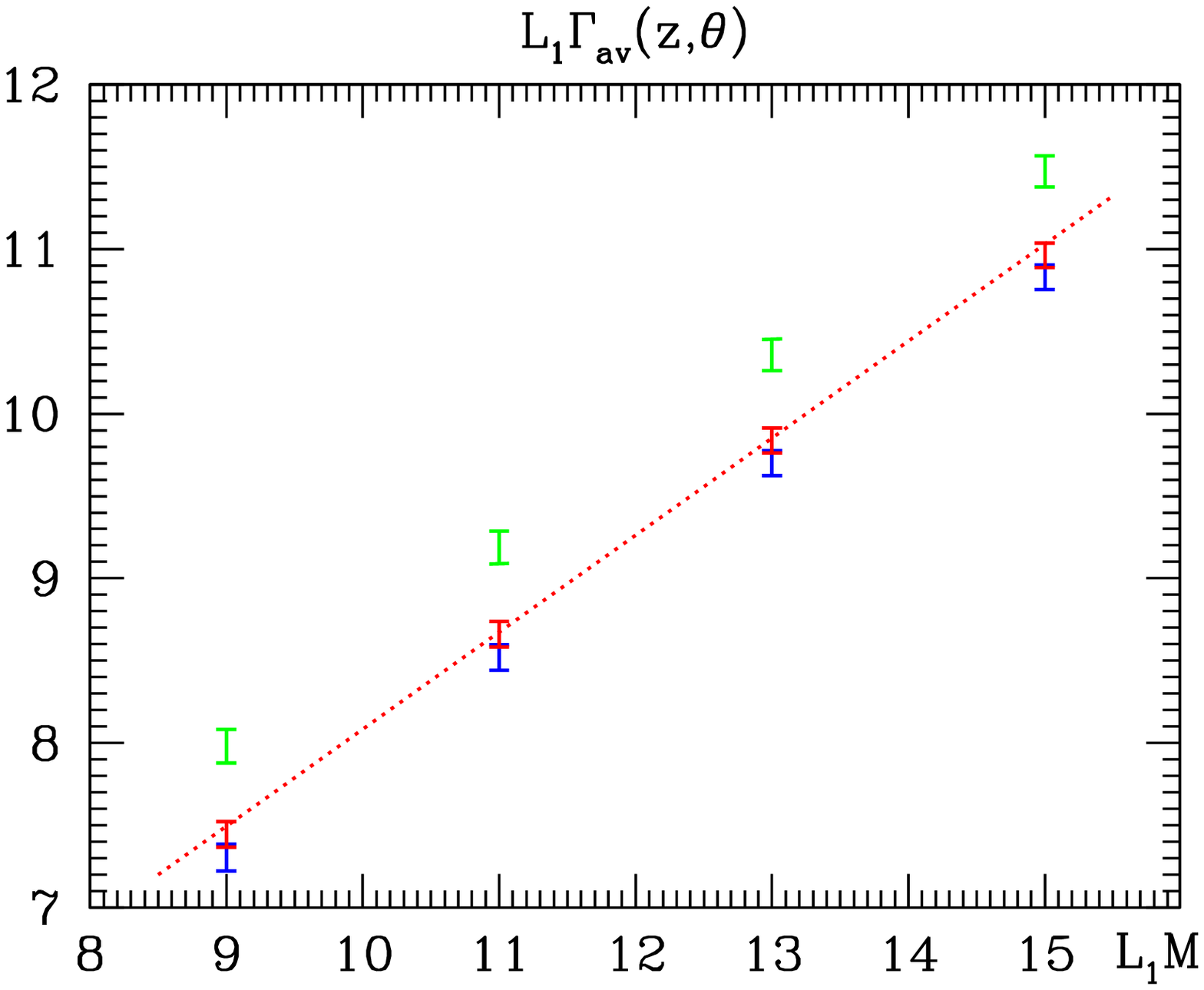,width=0.23075\textwidth}
\vspace{-1.375cm}
\caption{\sl%
Preliminary continuum limit (left) and $z$--dependence (right) of the
spin-averaged B-meson energy in finite-volume QCD for $\nf=2$.
}\label{fig:gamPSV_CL+zdep}
\vspace{-1.125cm}
\end{figure}

\Fig{fig:gamPSV_CL+zdep} shows an example for the mass dependence of a
finite-volume QCD observable in the continuum limit, which enters the 
matching step.
%
%
\section{Outlook}
\label{Sec_outl}
Thanks to technical and conceptual advances, most notably regarding
non-perturbative $\Or(a)$ improvement and renormalization in QCD and HQET, 
heavy-light physics has recently seen a significant progress.
Since most current simulations control other systematics such as dynamical 
quark effects and the continuum extrapolation, more precise results for 
the c- and b-quark masses can be expected within the next years.

\vspace{0.125cm}
\noindent {\bf Acknowledgments.}
I am indebted to my colleagues B.~Blossier, P.~Fritzsch, N.~Garron, 
G.~von Hippel, H.~Meyer, M.~Della Morte, S.~Sch\"afer, H.~Simma and 
R.~Sommer from the ALPHA Collaboration and to G.~De Divitiis, A.~J\"uttner 
and N.~Tantalo for an enjoyable, fruitful collaboration.
We thank NIC/DESY for allocating computer time on the APE and BlueGene/L\&P
computers to this project.
Support by the European Community through EU Contract 
No.~MRTN-CT-2006-035482, ``FLAVIAnet'' and by the DFG under grant
HE~4517/2-1 is acknowledged.

%
\bibliography{lattice_ALPHA}
\bibliographystyle{h-elsevier3.bst}
\end{document}